\documentstyle[twocolumn,aps,floats,graphicx]{revtex}

\begin{document}

\newcommand{\bec}{\begin{center}}
\newcommand{\ec}{\end{center}}
\newcommand{\be}{\begin{equation}}
\newcommand{\ee}{\end{equation}}
\newcommand{\beqn}{\begin{eqnarray}}
\newcommand{\eeqn}{\end{eqnarray}}
\newcommand{\bet}{\begin{table}}
\newcommand{\ent}{\end{table}}
\newcommand{\bib}{\bibitem}

\wideabs{

\title{
Possible three-dimensional chiral charge ordered superconducting state in cuprates
}

\author{P. S\"ule} 
  \address{Research Institute for Technical Physics and Material Science,\\
Konkoly Thege u. 29-33, Budapest, Hungary,\\
sule@mfa.kfki.hu
}

\date{\today}

\begin{abstract}

\maketitle

The 2D pair-condensate is characterized by a fluctuating chiral charge ordered state
with a "checkerboard" pattern in the $CuO_2$ planes and with an alternating supermodulation along the c-axis
in such a way that the adjacent layers are mirror images of one anothers electronic state.
Planar chiral order is revealed with a recent circular dichroism (CD) ARPES experiment (A. Kaminski, {\em et al.}, Nature, {\bf 416}, 611. (2002)).
We propose further CD experiments on ultrathin films with varying thickness and argue that
the odd number of unit cells along the $c$-axis might provide dichroism hence might support the picture of 3D chirality in
cuprates.
We find that Coulomb energy gain occurs along the $c$-axis within a multilayer chiral charge ordered state, which is proportional to the measured bilayer condensation energy and to $T_c$ at optimal doping.
Within our approach the superconducting (SC) pair is composed of the hole content of the coherence area and the
self-repulsion of the condensate is compensated by the gain in the inter-layer Coulomb energy below $T_c$.
The SC condensate and the fluctuating charge order can also be described by a dynamical inter-layer electrostatic complementarity.
%The static $c$-axis dielectric constant $\epsilon_c$ and the coherence length $\xi_{ab}$ are also calculated for various cuprates
%and compared with the available experimental data.
\\
\vspace{0.2cm}

{\bf 74.20.Mn} Nonconventional mechanisms,
{\bf 74.20.-z} Theories and models of the superconducting state,
{\bf 74.72.-h} Cuprate superconductors,
{\small {\em keywords}: high temperature superconductivity, charge ordered state, checkerboard modulation,
inter-layer coupling scenario, Coulomb instability, chiral and magnetic ordering}\\
preprint available at {\em cond-mat/0312474}

\end{abstract}
}

\section{Introduction}

  A series of recent experiments explored the emergence of spin and charge order in cuprates
\cite{Tranquada,Mook,Lake,Khaykovich,Hoffman,Kivelson,Alff,Howald}.
Local probes such as scanning tunneling microscopy (STM) revealed recently evidence for a
static checkerboard (CB) charge pattern with a real-space modulation periodicity of $4a_0$ in the
vortex core of Bi2212 \cite{Hoffman}, which is a provocative evidence for pinned
charge stripes \cite{Kivelson}.
A $4a_0 \times 4a_0$ charge pattern has also been reported in the absence of the applied magnetic field
\cite{Howald}, which is possibly induced by impurities at the surface.
The STM observations could be consistent with a dynamical fluctuation of charge order states (COSs) which is expected to slow down
with the increasing magnetic field and a static charge ordered pattern emerged in and around the vortex core \cite{Chen}.
It has been suggested that 
superconductivity and magnetic charge order coexist at moderate magnetic field and spin density
wave order grows with the applied magnetic field \cite{Khaykovich}.
Static magnetic charge order has also been observed in $YBa_2Cu_3O_{7-\delta}$ (YBCO) by neutron scattering experiment
without the need of impurity doping and seems to be compatible
with superconductivity as long as spin order remains dynamic \cite{Mook}.
 The ground state of cuprate superconductors is also characterized by the interplay between competing and coexisting
ground states \cite{Alff}.
A 4-unit-cell superstructure ($4a_0 \times 4a_0$) is also reported in a high-energy X-ray diffraction study in optimally doped
YBCO \cite{Islam}.

  The CB COS of cuprates attracted recently the attention of several theoreticians as well
\cite{Chen,Alexandrov,Zhu}.
It has been proposed that the magnetic field destroys the phase coherence of the hole pair by localizing them
into a crystal, without breaking the pair \cite{Chen}.
In particular the CB modulated superconducting (SC)
COS has been found favorable in an intermediate doping interval using the t-J model \cite{Vojta}.
 It is still a question whether the modulations are present only in the local density of electronic states (LDOS)
or also in the charge density \cite{Sachdev,Podolsky}.  
If the SC COS can be described by a weak CB modulation \cite{Sachdev},  
little modulations of the charge density may be hard to detect in the integrated LDOS \cite{Hoffman} hence
the improvement of the experimental resolution will be important in the future.

  Following a recent communication \cite{Sule_JS}, in this work, we propose a specific microscopic state for the SC state in cuprates.
This state has a $4a_0 \times 4a_0$ CB symmetry as observed in the experiment together with
a local chiral symmetry (inter-layer mirror symmetry).
A phase transition to this kind of a state where two-dimensional (2D) parity ($P$) and time-reversal symmetries ($T$) are simultaneously
violated has been proposed and studied in a number of articles \cite{Laughlin,Chakravarty}.
Up to date only planar
chiral order is found \cite{Kaminski} and early circular dichroism (CD) experiments have lead to controversial results
in the observation of three-dimensional (3D) chirality in cuprates \cite{Lawrence,Lyons}.
We propose to recall these experiments and suggest further measurements which might support
the existence of a possibly hidden (local) electronic 3D chiral order in the superconducting state based on a number of a theoretical reasonings and
experimental findings (section IV).

  We propose a simple phenomenological model for explaining HTSC
in cuprates in which these materials are nearly in a real space pairing limit \cite{Carlson}.
The pairs of holes form two particle bound states, and then Bose condense at $T_c$ with the spatial
constraint that the number of pairs per coherence area is approximately equal to $1$. 
In order to avoid the frustration caused by the propagation of hole pairs \cite{Carlson} we propose
a "scattered" density wave model of the hole pair state, in which the charge density of the holes are distributed in the
coherence area.
In this sense the CB pattern might be an interference pattern due to elastic quasiparticle (QP) scattering on the Fermi surface of the superconductor \cite{Hoffman}.
Inter-layer (IL) Coulomb energy gain may arise due to the occurance of
an alternating (asymmetric) interference charge pattern in the adjacent layers.
In this case the sign of the broken $P$ and $T$ symmetry alternates between planes and the broken symmetry
may be detected in principle by surface-sensitive probes or by weak effects in neutron scattering \cite{Halperin}.

  We would like to show that instead of using a complicated quantum formalism, it is possible to understand
certain properties of HTSC dealing only with a classical picture of inter-layer electrostatics of cuprates.
 Although for simplicity we present a static charge order, the real charge order is dynamic in the
SC state (with static spin density wave order).
The static charge order might only appear upon high external magnetic field (together with dynamic spin order), which breaks, however down
SC due to the pinning of holes to the lattice (COS of the Wigner solid).
Nevertheless, we model here the SC state by a static COS which can be taken as a "snapshot" of the
dynamical picture. The resonance of these pictures leads to the realistic picture of the dynamical charge order
in the SC state (charge fluctuations).

  Furthermore we assume the alternation (supermodulation) of the "checkerboard" charge pattern along the c-axis (that is normal
to the planes) which leads to Coulomb energy gain (see FIG 1). We will discuss the possible experimental support of this assumption (see section IV). Without this assumption inter-layer {\em Coulomb instability}
(CI) occurs in layered cuprates due to the enormous IL repulsion of holes.
At best of our knowledge there is no widely accepted theory for avoiding the enormous IL Coulomb repulsion induced
by the holes, although the role of holes in superconductivity is essential.
In any naive model of electron pairing in cuprates the in-plane  Coulomb repulsion is also troublesome.
When the pairs of charge carriers are confined to the sheets, naturally a net self-repulsion of the pair condensate
occurs. Although short range Coulomb screening in the dielectric crystal can reduce the magnitude of the
repulsion it is insufficient to cancel Coulomb repulsion completely \cite{Freedberg}.
{\em We would like to show that using a simple CB COS picture one can avoid both the in-plane and 
the out-of-plane CI of the layered SC condensate}.

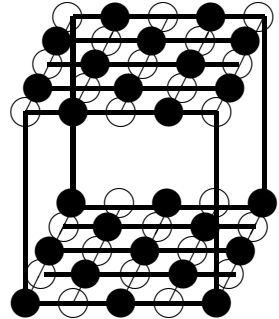
\begin{figure}

\setlength{\unitlength}{0.05in}
\begin{picture}(30,30)(-22,4)
%\thicklines
\linethickness{0.55mm}
  \put(0,0){\line(1,2){5.0}}
  \put(20,0){\line(1,2){5.0}}
  \put(0,20){\line(1,2){5.0}}
  \put(20,20){\line(1,2){5.0}}
  \put(5,0){\line(1,2){5.0}}
  \put(10,0){\line(1,2){5.0}}
  \put(15,0){\line(1,2){5.0}}
  \put(5,20){\line(1,2){5.0}}
  \put(10,20){\line(1,2){5.0}}
  \put(15,20){\line(1,2){5.0}}

  \put(20,0){\line(0,1){20.0}}
  \put(0,0){\line(0,1){20.0}}
  \put(0,20){\line(1,0){20.0}}
  \put(0,0){\line(1,0){20.0}}
  \put(5,30){\line(1,0){20.0}}
  \put(5,10){\line(0,1){20.0}}
  \put(25,10){\line(0,1){20.0}}
  \put(5,10){\line(1,0){20.0}}
  \put(2,3){\line(1,0){20.0}}
  \put(2.5,5.5){\line(1,0){20.0}}
  \put(4,8.){\line(1,0){20.0}}
  \put(1.3,22.5){\line(1,0){20.0}}
  \put(2.3,25.0){\line(1,0){20.0}}
  \put(3.5,27.5){\line(1,0){20.0}}
%  \multiput(0,0)(0,5){5}{\line(1,0){20}}
%  \multiput(0,0)(10,0){5}{\circle*{3}}
  \put(0,0){\circle*{3}}
  \put(5,0){\circle{3}}
  \put(10,0){\circle*{3}}
  \put(15,0){\circle{3}}
  \put(20,0){\circle*{3}}
%  \put(15,15){\circle*{8}}
  \put(0,20){\circle{3}}
  \put(5,20){\circle*{3}}
  \put(10,20){\circle{3}}
  \put(15,20){\circle*{3}}
  \put(20,20){\circle{3}}
  \put(1.6,3){\circle{3}}
  \put(6.3,3){\circle*{3}}
  \put(11.5,3){\circle{3}}
  \put(16.5,3){\circle*{3}}
  \put(21.3,3){\circle{3}}
  \put(2.5,5.5){\circle*{3}}
  \put(7.5,5.5){\circle{3}}
  \put(12.5,5.5){\circle*{3}}
  \put(17.5,5.5){\circle{3}}
  \put(22.5,5.5){\circle*{3}}
  \put(4,8){\circle{3}}
  \put(8.6,8){\circle*{3}}
  \put(13.8,8){\circle{3}}
  \put(18.8,8){\circle*{3}}
  \put(23.8,8){\circle{3}}
  \put(4.8,10.5){\circle*{3}}
  \put(9.8,10.5){\circle{3}}
  \put(14.8,10.5){\circle*{3}}
  \put(19.8,10.5){\circle{3}}
  \put(24.8,10.5){\circle*{3}}
  \put(1.3,22.5){\circle*{3}}
  \put(6.3,22.5){\circle{3}}
  \put(10.8,22.5){\circle*{3}}
  \put(16.4,22.5){\circle{3}}
  \put(21.4,22.5){\circle*{3}}
  \put(2.3,25.0){\circle{3}}
  \put(7.3,25.0){\circle*{3}}
  \put(12.3,25.0){\circle{3}}
  \put(17.3,25.0){\circle*{3}}
  \put(22.2,25){\circle{3}}
  \put(3.3,27.5){\circle*{3}}
  \put(8.5,27.5){\circle{3}}
  \put(13.2,27.5){\circle*{3}}
  \put(18.3,27.5){\circle{3}}
  \put(23,27.5){\circle*{3}}
  \put(4.4,30.0){\circle{3}}
  \put(9.4,30.0){\circle*{3}}
  \put(14.4,30.0){\circle{3}}
  \put(19.4,30.0){\circle*{3}}
  \put(24.4,30.0){\circle{3}}
\end{picture}

\vspace{1cm}
\caption{\small The alternating chiral "checkerboard" charge ordered state of the bilayer hole-antihole condensate mirror images
in the $4a_0 \times 4a_0$ bilayer superlattice model.
Only a bilayer is plotted, however, this bilayer pattern alternates along the c-axis as well as
the pattern is periodic along the ab-plane.
Each lattice sites (opened (holes) and filled (antiholes) circles) correspond to a $CuO_2$ unit cell. 
Note the charge asymmetry between the adjacent layers. The bilayer can accommodate a pair of a boson condensate ($4e$).
The inter-layer charge complementarity of the charge ordered state is crucial for getting inter-layer Coulomb
energy gain.
}

\end{figure}

\section{Charge distribution in the checkerboard charge order state}

   The CB COS can be understood as the in-plane alternation of hole-electron pairs in such a way that the
Cooper wave-function is "composed" of the hole content $p$ of the coherence area. The lattice site centered partial charges of the QP charge density is denoted as antihole partial charges ($q_{ah}$). 
The sum of antihole charges at various lattice sites correspond to the $2e$ charge of the quasiparticle (QP)
within the coherence area (charge sum rule: $\sum_i q_{ah,i} \approx 2e$, where $q_{ah,i}$ is the partial antihole charge at lattice site $i$).
Therefore within our representation of HTSC the QP charge density is shared by each $CuO_2$ sites in the
coherence area.
This is a marked difference with other theories (e.g. RVB \cite{PWA}) where the SC pairs are formed 
between electrons of neighboring sites. In these theories the entire coherence area does not
take part in the formation of QPs and hence the problem of self-Coulomb repulsion is huge.

  In the CB
representation of the SC state every second lattice site ($CuO_2$ sites) is occupied by antiholes (the scattered nodes of the QP) and the rest are holes with partial charges of $q_h$. 
$25$ $CuO_2$ unit cells can be found in the $4a_0 \times 4a_0$ CB, and $25 \times p_0 \approx 4e$ hole content
is provided by the coherence area at optimal doping, where $p_0 \approx 0.16e$ is the optimal hole concentration found in various
single- and double-layer cuprates \cite{Presland} which are described by the empirical relation
$T_c/T_c^{max}=1-82.6(p-0.16)^2$ \cite{Presland} where $p$, the hole content varies in a broad range.
$p_0 \approx 0.16e$ is also found by first principles calculations as an optimal
hole concentration \cite{Sule}.
Although $p$ is measured in the normal state (NS) of cuprates, one can assume, based on
simple chemical and physical intuity, that the chemical affinity of the $CuO_2$ unit might be similar in the
SC and in the NS of cuprates. Therefore, if $p \approx 0.16e$ in the NS, similar
quantity can also be assumed for the condensate antihole charge/$CuO_2$ units.
For simplicity we consider only optimal doping, extension of this speculation is straightforward
for the entire doping regime by varying $q_{ah,i}$ expecting a correspondence between $p$ and the 
QP partial charges in the SC state.

  When the hole content
condenses below $T_c$, $2e$ charge is used for the charge neutralization of every second hole sites leading
to an antiferromagnetic (AF) background COS possibly also with a CB modulation.
The SC pair is formed from 
the other $2e$ charge coming from the $4e$ hole content on the AF background.
The reason for the separation of the $4e$ hole content of the coherence area into $2e$ charge neutralizing charge and
to $2e$ QP pair is a somewhat hypothetical, however seems to be a reasonable choice.
By virtue only a single QP pair can be localized within the coherence area and therefore the other $2e$ charge
is strongly coupled (pinned) to the lattice forming a delocalized AF background. 

  The QP charge condenses to the charge neutralized sites leaving there antihole excess partial charges.
The partly charge neutralized hole doped sheet possesses also a CB pattern with an antiferromagnetic (AF) spin order. 
The QP $2e$ charge condenses to the neutralized sites forming the antiholes in this AF background  and a SC CB charge pattern occurs.
%otherwise an undoped AF insulating state would be emerged.
The excess charge density at the antihole sites leads to the increase of the on-site repulsion.
In order to maintain the phase coherence of the QP nodes (antiholes), the pairing-glue is necessary
to compensate the effect of the increased on-site repulsion.
We argue in the rest of the article that the pairing-glue is provided by the gain in the IL Coulomb energy
and which supports pairing up to $T_c$ where the thermal fluctuation breaks the phase coherence of the QP pair.

 Charge fluctuations (CF) might occur due to the hopping of the antiholes to the holes (hole annihilation).
CF is a {\em collective phenomenon} in the AF background (the coherent motion of holes), the antiholes hop from an antihole site to a hole simultaneously keeping
the phase coherence of the $2e$ pair and leading to a fluctuating order \cite{Kivelson}. 
The charge fluctuations are associated with an electronic liquid crystalline state (melted Wigner crystal)
\cite{Sachdev}.
This kind of a dynamic picture of the charge density wave order might be consistent with the presence of
a periodic modulation in the electron hopping or pairing amplitude \cite{Podolsky}.
Therefore the collective hole-antihole "exchange" between neighboring lattice sites leads to a resonance between CB COSs.
Within our picture the motion of a hole is accomplished by the coherent motion of partial holes (the nodes of the scattered hole pair) as well as
the motion of a QP is associated with the coherent motion of partial charges (antiholes, the scattered nodes of the condensate charge density waves) in the liquid condensate.
The site-by-site motion of a complete hole (a complete missing electron at a site) is an unfavorable process with a decent barrier 
and leads to an energetically frustrated system \cite{Carlson}.
Instead we propose the motion of the scattered density waves or holes on the lattice.

   Distributing the $2e$ charge of the quasiparticle condensate in the $4a_0 \times 4a_0$ coherence area the black and
white "fields" of the CB correspond to lattice site centered antiholes and holes with partial charges of
$q_{h,ah} \approx 4e/N^2 \approx \pm 0.16e$, where $N$ is the real space periodicity of the coherence
area ($N \approx \xi_{ab}/a_0+1, N=5$ for the $4a_0 \times 4a_0$  lattice, $N^2$ is the number of $CuO_2$ unit cells in the coherence area) where $\xi_{ab} \approx 16 \hbox{\AA}$ is
the in-plane coherence length \cite{Tinkham} and $a_0 \approx 3.9 \hbox{\AA}$ is the lattice constant.

  We would like to calculate the IL electrostatic energy of an alternating multilayered CB COS, which is
essential to avoid Coulomb instability even in the vortex core.
The multilayered CB COS might be energetically competitive with other possible non-alternating pattern in the SC state (e.g. stripes).
In the CB state the condensate does not "suffer" from the in-plane and out-of-plane CI, and also energy gain occurs
due to the out-of-plane alternation of holes.
Within the planes the net Coulomb self-interaction nearly vanishes due to the equal number of repulsive and
attractive pair interactions (inter-site) in the CB alternation of holes and antiholes.
In other words, the planes are sufficiently relaxed from Coulomb forces (electrostatically balanced), otherwise the 2D structure could not be retained.
Indeed, in most of the cuprates no out-of-plane distortion of the $CuO_2$ layers (buckling) is observed, except in
YBCO. Even in this case buckling is weak and is observed in the NS \cite{Plakida}.
Other possible COSs (e.g. stripes) with no CB pattern have a tendency to be electrostatically unbalanced
and, therefore, energetically unfavorable.
The only source of the energy gain in the potential energy is provided by the IL Coulomb coupling which is converted to the condensation
energy of the SC state (the free energy gain when the system goes to the SC state).
%------------------------------------------------------
\begin{figure}[hbtp]
%\begin{figure}[!t]
\begin{center} 
\includegraphics*[height=5.5cm,width=7.5cm]{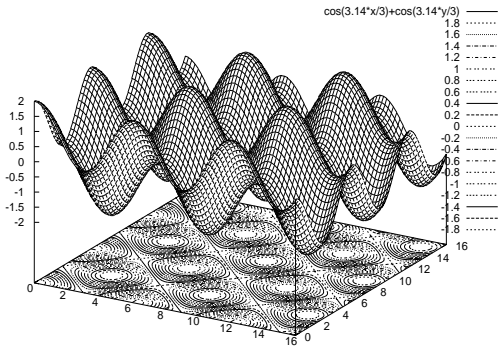}
\caption[]{
The 3D view of
the static "checkerboard" charge pattern of the boson condensate
corresponding to the $4a_0 \times 4a_0$ charge modulation represented
by the order parameter given in Eq. (1).
$x$ and $y$ coordinates are given in $\hbox{\AA}$.
The wells and peaks correspond to holes and antiholes, respectively.
}
\label{cond_tc}
\end{center}
\end{figure}
%------------------------------------------------------

  The kinetic energy driven mechanism of HTSC is also proposed as a possible source of pairing \cite{Carlson,Molegraaf}. 
It must be admitted that we can not rule out the kinetic energy gain mechanism and more sophisticated
approaches will be necessary to get a clear cut conclusion on this problem.
We simply remark here that owing to the law of the conservation of energy the gain in the potential energy side must
be converted to an increase in energy on the kinetic side and vice versa.  
This basic law implies that even if saving in kinetic energy is responsible for HTSC then there must be an energy
loss on the potential energy side.
The advocates of the kinetic energy mechanism attribute the potential energy loss naturally to the pairing induced increase
in the planar repulsion energy \cite{Carlson}.
In this paper we would like to show that it is possible to account for $T_c$ and for the condensation energy using
only the charge complementarity induced IL Coulomb energy gain.

 We would like to study then the magnitude of direct Coulomb interaction
between charge ordered square superlattice layers as a possible source of pairing interaction.
Our intention is to understand HTSC within the context of an IL Coulomb-mediated mechanism.
 The IL charging energy we wish to calculate depends on the IL spacing, the IL dielectric
constant $\epsilon_c$, the hole content $p$ and the size of the superlattice ($\xi_{ab}$).
Finally we calculate the static $c$-axis dielectric constant $\epsilon_c$ and the coherence length $\xi_{ab}$ for various cuprates
which are compared with the experimental observations.

\section{The superconducting order parameter}

  The {\em superconducting order parameter} (OP) of the d-wave condensate which corresponds to the model with a checkerboard charge modulation in the planes takes the form of
\be
 \Psi(x,y)=n_0^{1/2} [cos(\frac{x}{a_0} \pi)+cos(\frac{y}{a_0} \pi)].
\label{OP}
\ee
For simplicity, the distribution of the OP is neglected in the 3rd dimension and a nearly perfect
2D character is attributed to the condensate.
The 3D anisotropy of the condensate is negligible in the superconducting (SC) state 
which is reflected by the ratio of the in-plane and out-of-plane coherence lengths $\xi_{ab}/\xi_c \approx 10$ \cite{Tinkham}.
The factor $n_0$ is the maximal value of the charge density at the lattice site centers.
Eq.~(\ref{OP}) is displayed for the coherence area in FIG 2.
The modulation of the order parameter corresponds to the real-space modulation of the
hole density in the superconducting (SC) state.
This kind of an order parameter is given earlier by Alexandrov \cite{Alexandrov}.

 The order parameter must satisfy the charge sum rule for the boson condensate 
indicating the localization of the pair condensate within the coherence area,
\be
 2 \approx \int_0^{\xi_{ab}} \vert \Psi(x,y) \vert^2 dx dy.
\label{OPsumrule}
\ee
Another restriction on $\Psi(x,y)$ is that its integral over 
a unit cell with the area of $\sim (a_0/2)^2$ must correspond to the partial point charges of
holes and antiholes in the CB COS, 
\be
 \vert q_{h(ah)} \vert=\int_0^{a_0/2} \vert \Psi(x,y) \vert^2 dx dy.
\label{sumrulehole}
\ee
$q_{h(ah)}$ can also be taken as a free parameter ("filling" factor) at the lattice sites.
We assume homogeneous filling, all the lattice sites occupied by the same $q_{ah}$ or
$q_h$.
Calculations indicate
that the magnitude of the calculated condensation energy is in accordance with the experiment
when $q_{h(ah)} \approx p \approx 0.16e$. The variation of $q_{h(ah)}$ introduces the effect of doping.
This equation reflects the lattice site centered localization of the holes and antiholes
and leads to a simple electrostatic model where the charge modulation for simplicity is replaced
by classical point charges centered in the center of the holes and antiholes.
However, it must be admitted that in certain cuprates, usually with low $T_c$, the coherence area is
larger then $4a_0 \times 4a_0$. In these cases $q_{h(ah)} \ne 0.16e$ at optimal doping.
For instance in LSCO ($La_{1.85}Sr_{0.15}CuO_{4+\delta}$) $\xi_{ab} \approx 7a_0$, therefore $q_{h(ah)}=4e/64 \approx 0.063e$ at optimal doping
which leads to a weak modulation both in the LDOS and in the charge density representing a great
challenge for the experiment.

One can also speculate on the exponential decay of the OP at the border of the coherence area leading to
inhomogeneous distribution of $q_{h(ah)}$ within the CB similar to that of in refs. \cite{Chen,Zhu}.
For simplicity we are dealing only with the homogeneous distribution of $q_{h(ah)}$ within the CB which 
can be taken as an average distribution of the charge pattern. Charge inhomogeneity
could also be troublesome since partial charges $\vert q_{h(ah)} \vert > 0.16$ 
could be energetically unfavorable due to the increasing on-site repulsion with increasing charge density
at certain regions of the CB.

  IL Coulomb energy gain occurs only in that case when 
holes in one of the layers are in proximity with antiholes in the other layer
(FIG 1, IL {\em electrostatic complementarity}, bilayer model).
An important feature is then that
the boson condensate can be described by an IL {\em charge asymmetry} (mirror symmetry).
Therefore we assume an alternating charge pattern along the c-axis.
The IL coupling of the boson-boson pairs in the bilayer $5 \times 5$ model naturally suggests the
effective mass of charge carriers $m^{*} \approx 4 m_e$, as it was found by measurements \cite{krusin}.
Furthermore, the CFs in different planes are also coupled, in order to keep charge complementarity 
even during the fluctuation ("superexchange") of the CB charge patterns.

 Important to note that the charge sume rule
holds for the characteristic bilayer with a coherence area $\sum_i^{N^2} q_i^{ahole}=4e$.
In other words a pair of a boson condensate can be localized within a characteristic bilayer depicted in FIG 1.
 The alternation of the charge pattern along the c-axis is reflected in the order parameter
by the layer-by-layer alternation of $cos$ and $sin$ functions 
as well as the in-plane CF might also be reflected by the "superexchange" (coherent self-exchange) of $cos$ and $sin$ functions due to the collective motion of holes and antiholes.
This is nothing else then the manifestation of the sign reversal of the order parameter which is seen
by tunneling experiment as well \cite{Wollman}.

\section{Possible experimental fingerprints of the chiral order}

  Although a direct experimental evidence is not reported until now for the $c$-axis alternation of the 
hole density we discuss certain measurements which might be related to the presence of a chiral order
in cuprates.
The experimental detection of the electronic chiral order could be difficult in the geometrically achiral cuprates 
(due to the lack of the geometric inversion center, noncentrosymmetric compounds) therefore we discuss various measurements
to get a comprehensive picture.

 The recently published {\em circular dichroism ARPES} experiment \cite{Kaminski} is consistent directly
with a planar chiral order.
The experimental setup in these experiments incorporates a mirror plane perpendicular to the
$ab$-planes hence the chirality of the ab-plane is explored ($(0,0)-(\pi,\pi) (\Gamma-M$ line)) \cite{Kaminski_com}. 
This measurement provides a provocative evidence for a broken symmetry state below $T^{\star}$,
where a pseudogap appears in the underdoped Bi2212 compound.
Originally it was proposed by Varma {\em et al.} to support the circular orbital current model \cite{Varma}
which might be consistent with a circular charge motion picture within a CB pattern with a coherence area.
According to our proposal with IL coherence of charge motion the circular currents are 
must be in the same direction with each other in the adjacent layers to observe dichroism also along
the $c$-axis.
Circular currents with opposite orientation in the adjacent layers would not lead to dichroism because the
adjacent layers form an enantiomorphous pair with globally restored symmetry hence no difference in absorption between right and left
circularly polarized radiations could be found.

  Unfortunately an experimental test of 3D chirality is difficult to make directly.
It is hard if impossible to repeat the CD ARPES measurements
with a mirror plane parallel to the $ab$-plane owing to the fact that thin films are in general use
in ARPES measurements \cite{Kaminski_com}.
Another problem is that in many AF crystals, where the locally time-reversal symmetry can be
globally restored, the total dichroic signal is zero \cite{Matteo}.
The failure of observing circular dichroism in cuprates \cite{Lawrence} could be 
attributed then to the macroscopic symmetry of these systems (hidden chiral order), for example to
the formation of an optically inactive phase (a racemate with enantiomorphous pairs of layers).
Globally no dichroism occurs also in the $c$-axis alternating CB COS superstructure because the
crystal is stacked by enantiomorphous pairs of layers (also without the assumption of circulating currents). 
We place in this case the inversion center to the midpoint of the coherence bilayer shown in FIG 1.

  {\em Ultrathin films with odd number of layers, however, should provide nonvanishing dichroic signal (optical rotation) because of
the presence of an unpaired optically active layer.} 
In ultrathin layers the number of layers can be kept under control, even one-unit-cell thick sample
can be prepared \cite{Li}.
Its signal is not compensated by a signal of its mirror image hence should be detected by CD measurements.
Therefore, we propose to repeat the early CD measurements with odd number of layers, hence double-layer
cuprates (such as YBCO or Bi2212) are unsuitable for the detection of 3D chirality with traditional
CD techniques. 
The CD measurements on thin layer YBCO \cite{Lawrence} therefore naturally does not result in 3D chirality.
Instead the ultrathin film of the single-layer LSCO should be used with odd number of unit cells along
the $c$-axis.
Although Lyons {\em et al.} \cite{Lyons} found no sign of optical activity in LSCO, 
the application of specific conditions proposed above might
indicate optical activity in the under- or optimally doped LSCO. 
Ultrathin films with varying thickness made from the three-layer Bi1223 compound should also be suitable 
for CD studies.
Within our model a one-unit-cell thick single-layers do not show HTSC due to the lack of IL coupling. 
Gelfand and Halperin \cite{Halperin2} have also proposed the refinement of optical experiments
with odd number of layers which would reveal clear evidence of broken time-reversal symmetry and of
anyons in cuprates.
At best of our knowledge since then no experiment has been done using such conditions.

 Moreover, the dichroism signal should alternate with the parity of the number of layers in the thin films:
the even number of layers should give vanishing signal while the odd number case should result in a decent
absorption difference between the right and left
circularly polarized radiations.
It should be noted that we do not question the remark of Lawrence {\em et al.} \cite{Lawrence} that 
the strong temperature dependence of the dichroic signal found in YBCO and in bismuthates \cite{Lyons}
must be attributed to surface inhomogeneities and to contaminations. 
We simply propose to apply these experiments for other compounds with the systematical variation of the film thickness
together with the careful treatment of the surface.
The strong CD signal below $T_c$ found in the single-layer bismuthate ($K_{1-x}Rb_xBiO_3$) \cite{Lyons} according to our knowledge is 
never falsified.

 In any case we also propose to probe the chiral order in the slightly underdoped cuprates in the
presence of a strong magnetic field where a quasi static charge order emerged possibly with
an internal asymmetry. In the SC state, however, the fluctuating magnetic order might weaken the signals coming from
the presence of the chiral order hence the detection of the charge asymmetry could be difficult.

  Recently the anomalous X-ray diffraction has been proposed to circumvent the limitation of the global symmetry, where
the local transition amplitudes are added with a phase factor that can compensate
the vanishing effect due to the global symmetry \cite{Matteo}.
Unfortunately no application of X-ray dichroism is known for cuprates until now.
  Optical activity has been detected using the natural circular dichroism in quite a
few noncentrosymmetric crystals \cite{Goulon}.
The x-ray magnetochiral dichroism has also been proposed
to unravel hidden space-time symmetry properties of magnetoelectric crystals \cite{Goulon}.
The application of these techniques for cuprates might also reveal a possible hidden chiral order
in the SC state.
  At present the experimental test of a chiral COS in SC cuprates remains, however, indirect.

 There are several indirect experimental findings which support the IL scenario, such as the $T_c$ variation
in thin film heterostructures \cite{Li,Goodrich}. In such films HTSC vanishes when the film thickness approaches
the IL distance.  
The variation of $T_c$ upon the external hydrostatic pressure in Hg-cuprates compounds \cite{Gao}
can also be explained by the IL coupling concept, since upon pressure the IL spacing of cuprates decreases
seriously which leads to an increase in $T_c$ due to the possible enhancement of IL coupling.

 The ARPES circular dichroism experiment reveals that the breaking of the time-reversal symmetry appears in the underdoped
samples while is not present in the overdoped regime \cite{Kaminski}.
The absence of planar chirality in the overdoped samples might be consistent with a 2D $\leftrightarrow$ 3D charge transfer of the hole content \cite{Sule_JS} which does
not occur in the underdoped regime when the planes remain coupled hence mirror symmetry is retained.
In the overdoped side, however, the strong 2D $\leftrightarrow$ 3D charge fluctuations screen IL coupling hence
IL symmetry progressively develops with overdoping.
Therefore
overdoping might destroy the chiral order (decouples the layers) and favors the occurrence of a symmetric Fermi-liquid phase.
Dispersion along the $\Gamma-Z$ line in ARPES spectra would be the indicative of 
2D $\leftrightarrow$ 3D CF although it is hard to collect such a data from presently
available ARPES results \cite{Damascelli}.

  The observed supression of the pseudogap in the overdoped side of the SC dome might also be associated with
the enhancement of 2D $\leftrightarrow$ 3D charge fluctuations. 
The pseudogap vanishes above the critical hole concentration $p \approx 0.19$ \cite{Tallon} which
might be associated with the decoupling of layers.
The superconducting peak ratio (SPR) seen in the ARPES spectra of Bi2212 peaks at $p \approx 0.19$
and drops for $p > 0.19$ \cite{Feng} which also support that IL decoupling effects become dominant
in the strongly overdoped regime.
This is because SPR can be related (not directly) to the superfluid density hence
the drop of the SPR signal intensity
reflects the reduction of the carrier concentration in the planes 
coexisting with the high doping level
and is associated with the 2D $\leftrightarrow$ 3D charge fluctuations
together with the lost of the 2D order in the strongly overdoped regime.
First principles calculations indicate the saturation of the in-plane hole concentration 
around $p \approx 0.19$ and support the presence of the 2D $\leftrightarrow$ 3D charge fluctuations
\cite{Sule}.
These calculations reveal that the hole concentration in the planes reaches its maximum
when the dopant band is completely occupied, e.g., for a closed O shell \cite{Sule}.
This occurs at around $p \approx 0.19$ and a strong $c$-axis anisotropy of the hole content
develops as demonstrated in the calculated LDOS \cite{Sule}
which also peaks at $p \approx 0.19$ \cite{Sule} in accordance with the specific heat anomaly \cite{Norman,Tallon}.
Interestingly the pseudogap line ($T^{*}$) crosses the SC dome at $p \approx 0.19$ \cite{Tallon}.
For $p > 0.19$ (strongly overdoped regime) $T^{*} \le k_B T_c$ and the hole content can freely
2D $\leftrightarrow$ 3D  fluctuate.
Thus we predict the occurrence of a 3D dispersion in the ARPES spectra along the out-of-plane
direction $\Gamma-Z$ in the Brillouin zone in the strongly overdoped regime and above $T_c$.
The strongest evidence for a coherent 3D transport is found in the overdoped Tl2201 
\cite{Hussey}.
The reported polar angular magnetoresistance oscillations in high magnetic field
firmly establishes the existence of a coherent 3D Fermi surface.
This is consistent with our 2D $\leftrightarrow$ 3D  fluctuation model of the hole content
in the heavily overdoped regime and in the normal state.

 The issue of bilayer splitting seen in ARPES spectra \cite{Norman,Damascelli} could also provide important ingredients
in searching for the signatures of chiral order and inter-layer coupling.
The two different component in the spectra is attributed to bonding and antibonding bands
with opposite symmetry along the c-axis with respect to the midpoint between the two $CuO_2$ layers
in the bilayer block of Bi2212 \cite{Damascelli}.
The overdoping induced appearance of a band with a strong c-axis anisotropy (due to 2D $\leftrightarrow$ 3D CF) could explain
bilayer splitting. This band might also provide a peak at ($\pi,0$) and could be the signature of
the broken chirality (the recovery of symmetry) in the normal state of cuprates.
This issue is still highly controversial and further specific studies are needed which provide
further informations on the 3D nature of HTSC.

\section{The condensation energy and $T_c$}

  We calculate the condensation energy/$CuO_2$ ($U_0$) in order to show that at a reasonable choice of the IL dielectric
constant $\epsilon_c$ the calculated $U_0$ is typically in the range of the measured values.
We will show in this section that this correlation between the measured and the calculated values manifest in a number of ways.
%------------------------------------------------------
\begin{figure}[hbtp]
%\begin{figure}[!t]
\begin{center}
\includegraphics*[height=4.5cm,width=6.5cm]{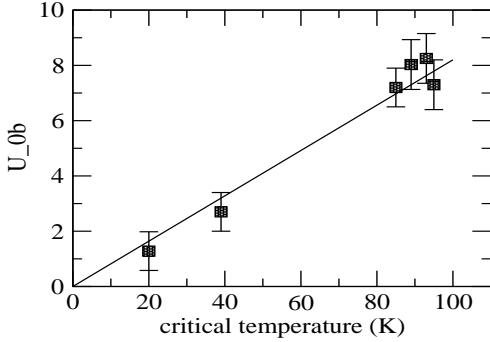}
\caption[]{
The bilayer condensation energy ($U_{0b}$, meV)
as a function of
the critical temperature (K) at optimal doping.
The straight line is a linear fit to the data. The slope of the linear fit
is $U_{0b}/T_c \approx k_B$ which is a strong evidence of Eq.~(\ref{kbtc}).
The error bars denote standard deviations estimated from various measurements of
the condensation energy.
}
\label{tc_cond}
\end{center}
\end{figure}
%------------------------------------------------------

 First the condensation energy of a bilayer with a coherence area in the planes ($U_{0b}$) is given as follows
\be
U_{0b}=2 (n+1) \biggm[ \frac{\xi_{ab}}{a_0}+1 \biggm]^2 U_0 \approx E_c^{IL,SC},
\label{gain_sc}
\ee
where $E_c^{IL,SC}$ is the Coulomb energy gain in the SC state.
$U_0$ is the experimental condensation energy given per unit cell. 
Eq.~(\ref{gain_sc}) is generalized for multilayer cuprates introducing
$n$.
For single layer cuprates $n=0$, for bilayers $n=1$, etc.
The factor $\biggm[ \frac{\xi_{ab}}{a_0}+1 \biggm]^2$ is the number of $CuO_2$ lattice sites
in the planes ($\approx N^2$, within the coherence area).
Factor $2$ is applied in Eq.~(\ref{gain_sc}) because we calculate the condensation
energy of a bilayer.
The IL Coulomb energy is in the SC state
\be
E_c^{IL,SC}=\frac{e^2 Q}{4 \pi \epsilon_0 \epsilon_c},
\label{IL}
\ee
where
\be
Q= \sum_{m=2}^{N_l} \sum
_{ij}^{N^2} \frac{q_i^{(n)} q_j^{(k)}}{r_{ij}^{(k,l)}},
%\label{IL}
\ee
where $r_{ij}^{(k,l)}$ is the inter-point charge distance and $r_{ij}^{(k,l)} \ge d_{IL}$, where $d_{IL}$ is
 the IL distance ($CuO_2$ plane to
plane, $i \ne j$). $k,l$ represent  
layer indexes ($k \ne l$).
$q_i^{(k)}$ and  $q_j^{(l)}$ are the point charges of holes and antiholes centered at $CuO_2$ lattice
sites in the $k$th and $l$th layers.
First the summation goes within the bilayer up to $N^2$ then the IL Coulomb interaction of the
basal bilayer are calculated with other layers
along the $c$-axis in both direction ($k=1,2$).
$N_l$ is the number of layers along the $c$-axis. When $N_l \rightarrow \infty$, bulk $E_c^{IL,SC}$ is calculated.

  The plot of $U_{0b}$ (the bilayer condensation energy) against $T_c$ is shown
in FIG 3 using only experimental data.
Remarkably the data points of various cuprates with a variety of critical temperature
fit to a line and its slope $U_{0b}/T_c$ is the Boltzmann constant $k_B$.
The average value we get is $k_B \approx 1.3 \pm 0.2 \times 10^{-23} J/K$ which is
remarkably close to the value of $k_B=1.38 \times 10^{-23} J/K$.
In the rest of the paper we will present further evidences in order to show
that the agreement might not be accidental.
According to the correlation found between $U_{0b}$ and $T_c$ 
the following formula can be given
using Eq.~(\ref{gain_sc}),

\be
U_{0b}=2 (n+1) \biggm[ \frac{\xi_{ab}}{a_0}+1 \biggm]^2 U_0 \approx k_B T_c.
\label{kbtc}
\ee
Therefore the bilayer condensation energy $U_{0b}$ can directly be related
to the thermal motion at $T_c$.
It must be emphasized that Eq.~(\ref{kbtc}) is coming directly from the empirical relation found
in FIG 3 and its theoretical derivation is currently not available yet.
This relation at a first look is certainly unusual and is completely different from the
formula known in conventional superconductors \cite{Tinkham}.

  We understand this correlation as follows: the SC pair-condensate is stable against
thermal fluctuations up to $T_c$, since the IL coupling energy $E_c^{IL,SC} \approx k_B T_c$.
Hence above $T_c$ the phase coherence of the QP pair is broken and SC is suppressed.
For $T > T_c$, in the underdoped side pinning of the holes becomes stable against CFs, whereas in the
overdoped side the hole content exhibits a strong 3D anisotropy hence the SC phase vanishes.
As far as the microscopic mechanism of pairing is concerned, we further argue that the
IL Coulomb energy gain is converted to the stabilization of the fluctuating CB COS, which is
although electrostatically is balanced (free of frustration), however, 
the increased on-site repulsion at the antihole sites (due to the increased density,
$q_{ah} \approx -0.16$) is energetically unfavorable and is compensated by the IL energy gain
up to $T_c$.
Above $T_c$ therefore the on-site repulsion of the condensate is no longer compensated
by the gain in the IL Coulomb energy.
The on-site repulsions of the QP charge density sum up to
the self-repulsion of the scattered QP pair which is compensated then by the out-of-plane
Coulomb energy gain (the pairing glue) below $T_c$.
The magnitude of the self-repulsion of the condensate with CB COS could easily be checked by sophisticated
{\em ab initio} or Hubbard-type approaches. Our rough estimate based on simple in-plane Coulomb energy calculations provides values in the range of $\sim k_B T_c$.
If the on-site repulsion of the condensate is compensated by the gain in the IL Coulomb energy then
it follows that the optimal hole content $p_0 \approx 0.16$ is more or less general for various cuprates
with different $T_c$ just because the coherence area increases with decreasing $T_c$ hence the on-site
repulsion is also decreasing with decreasing antihole partial charges (Table I).
Larger then $4a_0 \times 4a_0$ coherence area naturally suggests that $q_{ah} < p_0$, therefore
the remaining part of the hole content does not contribute to the QP charge density and
is rather pined to the lattice below $T_c$.
Above $T_c$ the phase coherence of the fluctuating CB COS is lost via hole annihilation
(slowed down CFs in the underdoped regime) leading to a frustrated AF spin order (spin glass) or via the $2D \leftrightarrow 3D$ phase transition of the hole content.

%%%%%%%%%%%%%%%%%%%%%
%%% TAB 1
%%%%%%%%%%%%%%%%%%%%
\begin{table}
%\center
\caption[]
{The calculated coherence length of the pair condensate
given in $a_0$ using the experimental condensation energies
of various cuprates and Eq.~(\ref{N}) at optimal doping.
}
{\scriptsize
\begin{tabular}{cccccc}
 & $T_c$ (K) & $k_B T_c$ (meV) & $U_0$ ($\mu eV/u.c.$)  & $\xi_{ab}^{calc} (a_0)$ & $\xi_{ab}^{exp} (a_0)$ \\ 
\hline
  Bi2201 & 20 & 1.6 & $10^a$  &  $\sim 8$ &       \\
 LSCO   & 39  & 2.5 & $21^b$ &  $\sim 7$ &  $5-8^c$  
 \\
 Tl2201 & 85 & 7 & $100 \pm 20^d$ & $\sim 5$ & \\
 Hg1201 & 95 & 7.8 & $80-107^e$ &  $\sim 5$ & $5^f$ \\
  YBCO  & 92 & 7.5 & $110^g$ &  $\sim 3$ & $ 3-4^h $ \\
  Bi2212 & 89 & 7.3 & $95^g$ &  $\sim 3-4$ & $4-6^i$\\
Hg1223  & 135 & 11 & 114     &  $\sim 4$   & $4^f$ \\
%---------------------------------------------------------------
\end{tabular}}
{\scriptsize
$a_0 \approx 3.88 \AA$,
$U_0$ is the measured condensation energy of various cuprates
in $\mu$ eV per unit cell at optimal doping.
$^a$ from \cite{MarelPC},
$^b$ $U_0 \approx 2$ J/mol from \cite{Loram,Momono},
$^c$ from \cite{Tinkham},
$^d$ \cite{Tsvetkov},
$^e$ $U_0 \approx 12-16$ mJ/g from \cite{Billon,Kirtley} and $\xi_{ab}$ from \cite{Thompson},
$^d$ $U_0 \approx 11$ J/mol from \cite{TallonLoram},
$^e$ $U_0 \approx 10$ J/mol from \cite{TallonLoram},
$^f$ from \cite{Thompson},
$^g$ from \cite{TallonLoram},
$^h$ from \cite{Tinkham},
$^i$ from recent measurements of Wang {\em et al.}, $\xi_{ab} \approx 23 \AA (\sim 5-6 a_0)$ \cite{Wang_sci},from STM images of ref. \cite{Hoffman} $\xi_{ab}\approx 4a_0$,
$\xi_{ab}^{calc}$ is calculated according to Eq.~(\ref{N}) and is also given in Table ~\ref{tab1} and $\xi_{ab}^{exp}$ is
the measured in-plane coherence length given in $a_0 \approx 3.9 \AA$.
The notations are as follows for the compounds:
Bi2201 is $Bi_2Sr_2CuO_{6+\delta}$,
LSCO ($La_{1.85}Sr_{0.15}CuO_{4+\delta}$),  
Tl2201 ($Tl_2Ba_2CuO_6$), 
Hg1201 ($HgBa_2CuO_{4+\delta}$),  
 YBCO ($YBa_2CuO_7$) and
 Bi2212 is $Bi_2Sr_2CaCu_2O_{8+\delta}$.
 Hg1223 is $HgBaCa_2Cu_3O_{12+\delta}$.
}
\label{tab1}
\end{table}

It is still also a possibility that only the phase coherence is broken and the in-plane fluctuation
of the holes persists (at least partly) above $T_c$ leading to the pseudogap phenomenon.
The incoherent motion of the charge carriers might also be then associated with 
the NS gap either within the planes and out-of-plane.
This issue certainly awaits further theoretical investigations.

 In order to test the validity of the empirical Eq.~(\ref{kbtc}) we estimate the 
coherence length of the pair condensate derived from Eq.~(\ref{kbtc})
and using only experimental data,
\be
\xi_{ab} \approx a_0 \biggm[ \sqrt{\frac{k_B T_c}{2 (n+1) U_0}}-1 \biggm]
\label{N}
\ee
The results are given in Table~\ref{tab1} as $\xi_{ab}^{calc}$ (in $a_0$ unit) and compared with the available
measured $\xi_{ab}^{exp}$. The agreement is excellent which strongly suggests that Eq.~(\ref{N})
should also work for other cuprates.
We predict for the multilayer Hg-cuprate Hg1223 ($T_c=135$ K) the condensation energy
$114$ $\mu$eV/u.c. and $\xi_{ab} \approx 4 a_0$ which is in agreement with the measured value \cite{Thompson}) using the dielectric constant of $\epsilon_c \approx 61$ for the Ca-spaced trilayer block.

 The expression Eq.~(\ref{kbtc}) leads to the very simple formula for the critical temperature
using Eq.~(\ref{gain_sc}) and a simple Coulomb expression for the IL coupling energy $E_c^{IL,SC}$
($T_c \approx k_B^{-1} E_c^{IL,SC}$)
\be
T_c (N,d,\epsilon_c) \approx \frac{e^2 Q}{4 \pi \epsilon_0 \epsilon_c k_B}  
\label{kbtc2}
\ee
When $N_l \rightarrow \infty$, bulk $T_c$ is calculated.
$T_c$ can also be calculated for thin films when $N_l$ is finite.
$\epsilon_c$ can also be derived
\be
\epsilon_c \approx \frac{e^2 Q}{4 \pi \epsilon_0 k_B T_c}  
\label{epsc}
\ee
where a $c$-axis average of $\epsilon_c$ is computed when $N_l \rightarrow \infty$.

%%%%%%%%%%%%%%%%%%%%%
%%% TAB 2
%%%%%%%%%%%%%%%%%%%%
\begin{table}
%\center
\caption[]
{The calculated dielectric constant $\epsilon_c$ using Eq.~(\ref{epsc})
in various cuprates at the calculated coherence length $\xi_{ab}$ of the charge ordered state
given in Table I.
}
{\scriptsize
\begin{tabular}{lccccc}
 & $d (\AA)$ & $T_c (K)$ & $\xi_{ab}^{calc} (a_0)$ & $\epsilon_c$ & $\epsilon_c^{exp}$  \\ \hline
 Bi2201 &    12.2 & 20   &   8  &  9.9 & 12$^a$ \\
 LSCO &      6.65 &  39 & 7 &  11.3 & $23 \pm 3, 13.5^b$  \\
 Hg1201    &     9.5   & 95  & 4 & 6.5 &    \\
 Tl2201 &    11.6 &  85 &  5 & 13.0 & 11.3$^c$ \\
  YBCO  &   8.5  & 93  &  3 & 19.4 & 23.6$^d$ \\ 
%---------------------------------------------------------------
\end{tabular}}
{\scriptsize
where $\xi_{ab}^{calc}$ is the estimated in-plane coherence length given in $a_0 \approx 3.9 \AA$ unit.
$d$ is the $CuO_2$ plane to plane inter-layer distance in $\AA$, $T_c$ is the experime
ntal critical temperature.
$\epsilon_c$ is from Eq.~(\ref{epsc}).
$\epsilon_c^{exp}$ are the measured values obtained from the following references: 
$^a$ \cite{Boris}, 
$^b$ \cite{epsilonc}, or from reflectivity measurements \cite{Tsvetkov},
$\omega_p \approx 55 cm^{-1}$ \cite{Sarma}, $\lambda_c \approx 3 \mu$m \cite{Kirtley},
$^c$ from \cite{Tsvetkov}, 
$^d$ from reflectivity measurements: $\omega_p \approx 60 cm^{-1}$ \cite{Sarma}, $\lambda_c \approx 0.9 \mu$m \cite{Kitano2}.
}
\label{tab2}
\end{table}

The calculation of the $c$-axis dielectric constants $\epsilon_c$ might provide
further evidences for Eq.~(\ref{kbtc}) when compared with the measured values \cite{Kitano,epsilonc}.
In Table ~\ref{tab2} we have calculated the static dielectric function $\epsilon_c$ using
Eq.~(\ref{epsc}) and compared with the experimental impedance measurements \cite{Cao,epsilonc}.
$\epsilon_c$ can also be extracted from the $c$-axis optical measurements using the
relation \cite{Tsvetkov}
$\epsilon_c(\omega) =\epsilon_c(\infty)-c^2/(\omega_p^2 \lambda_c^2)$, 
where $\epsilon_{\infty}$ and $\omega_p$ are the high-frequency dielectric constant and the plasma frequency, respectively \cite{Tamasaku}. $c$ and $\lambda_c$ are the speed of light
and the $c$-axis penetration depth. 
At zero crossing $\epsilon_c(\omega)=0$ and $\omega_p=c/(\lambda_c \epsilon_c^{1/2}(\infty))$.
Using this relation we predict for the single layer Hg1201 the low plasma frequency of
$\omega_p \approx 8$ cm$^{-1}$ using $\epsilon_c=\epsilon_c(\infty) \approx 6.5$ (Table II) and $\lambda_c \approx
8$ $\mu$m \cite{Kirtley} using $\xi_{ab} \approx 5a_0$ \cite{Thompson}.

\section{Conclusion}

  Our primary result is that the boson condensate in the superconducting state can be described
by an alternating "checkerboard" type of charge pattern along the c-axis which leads to 
inter-layer charge complementarity (chiral charge ordered state) and Coulomb energy gain. 
Our proposal is that this gain is converted to the condensation energy of the superconducting state, although the detailed mechanism of this process
still remains unclear. 
Within our model the pairing glue is also provided by inter-layer coupling.

  The superconducting state can also be described by a fluctuating order hence the phase coherence of
the quasiparticles can only be retained if inter-layer charge and phase complementarity emerged.
This physical picture naturally explains the variation of $T_c$ system-by-system in various
conditions (external and chemical pressure, multilayers, heterostructures etc.).
Without the assumption of inter-layer charge complementarity the superconducting state
would "suffer" from an enormous inter-layer Coulomb instability (repulsion) which is certainly not the case.

  The charge chirality model explains the recent circular dichroism ARPES experiment in which the
underdoped samples show dichroism below the characteristic temperature $T^{\star}$ of the pseudogap.
We believe that the modulation of the charge density along the $c$-axis theoretically provides the possibility of better understanding
HTSC.
  
  Finally we attribute the breakdown of SC on the overdoped side of the superconducting dome to the
destroy of the $2D$ order (the 2D $\rightarrow$ 3D transition of the hole-content in the normal state)
and to the emergence of an antiferromagnetic order on the underdoped side.
We support a picture which incorporates a fluctuating order in the optimally doped regime separated
from the Fermi-liquid-like overdoped regime by a quantum critical point and by the suppression of
charge fluctuations from the underdoped region.

\section{acknowledgment}

{\small
Special thank to E. Sherman for reading an earlier version of the manuscript
and for the helpful informations.
I would also like to thank for the helpful discussions with T. G. Kov\'acs,
I. Bozovic, A. Kaminski, N. Momono and for the stimulating comments to A. S. Alexandrov. 
I also acknowledge A. Damascelli for bringing to my attention ref. \cite{Hussey} and
for the useful discussions.

This work is supported by the OTKA grant F037710
from the Hungarian Academy of Sciences}
\\

%\newpage

\end{document}